\documentclass[referee]{aa}
\usepackage{psfig}

\newcommand{\be}{\begin{equation}}
\newcommand{\ee}{\end{equation}}

\begin{document}

\title{The very local Hubble flow: Simulating transition from chaos to order}

\author{
A. D. Chernin\inst{1,2} \and I. D. Karachentsev\inst{3} \and M. J.
Valtonen\inst{1,4} \and V. P. Dolgachev\inst{2} \and L. M.
Domozhilova\inst{2} \and D. I. Makarov\inst{3,5} }

\institute{Tuorla Observatory, Turku University, Piikki\"o, 21
500, Finland \and Sternberg Astronomical Institute, Moscow
University, Moscow, 119899, Russia \and Special Astrophysical
Observatory, Nizhnii Arkhys, 369167, Russia \and University of
West Indies, Trinidad and Tobago \and Isaac Newton Institute of
Chile, SAO Branch, Russia }

\authorrunning{A.D. Chernin et al.}
\titlerunning{The very local Hubble flow}

\date{Received / Accepted}

\abstract{The physical nature of the very local ($\le 3$ Mpc)
Hubble flow is studied on the basis of the recent high precision
observations in the Local Volume. A model including both
analytical treatment and computer simulations describes the flow
dynamical evolution from a chaotic Little Bang initial state to
the present-day state of a quasi-regular expansion. The two major
observational parameters of the flow, which are the rate of
expansion and the velocity dispersion, find a clear qualitative
and quantitative explanation in the model.

 }

\maketitle

\section{Introduction}

It has long been taken for granted in theoretical cosmology that
the notion of the cosmological expansion is applicable to only
very large spatial distances, and that only at the scale of great
clusters of galaxies, the markers that participate in the
cosmological expansion can be found. A reason for this comes from
the fact that the expansion with the linear velocity-distance
relation is directly associated with the uniformity of the
Universe, in the Friedmann standard cosmological theory. Since
the matter distribution is uniform only at the spatial scales
larger than 100--300 Mpc, the cosmological expansion is treated
as a phenomenon of the largest observed scales. However, in
drastic contradiction with this view, the phenomenon was
originally discovered by Hubble in the Local Volume within 20 Mpc
distance from us. The Local Volume is deep inside the cosmic cell
of matter uniformity, and observations reveal a significant
non-uniformity in the spatial distribution of galaxies there. How
may the regular linear velocity field be compatible with the
observed spatial non-uniformity of the galaxy distribution in the
Local Volume? How might the cosmological phenomenon be discovered
so closely to us?

The questions were clearly stated by Sandage (1986; see also
Sandage et al. 1972). In a recent paper, Sandage (1999) concludes
that an ``explanation of why the local expansion field is so
noiseless remains a mystery''. It is also puzzling that the local
rate of expansion is similar to the global one, if not exactly
the same, within 10--15 percent accuracy (Sandage 1999). This is
the Hubble-Sandage paradox described in more detail in our earlier
paper (Chernin, Karachentsev, Valtonen et al. 2004 --- hereafter
Paper I).

A possible solution to the Hubble-Sandage paradox has been
suggested by Chernin, Teerikorpi and Baryshev (2000) soon after
the discovery of cosmic vacuum (dark energy or the cosmological
constant) in distant supernova Ia observations (Riess et al.
1998, Perlmutter et al. 1999). It has been recognized (Chernin
2001, Baryshev, Chernin, Teerikorpi 2001) that cosmic vacuum with
its perfectly uniform density  makes the Universe effectively
uniform at various spatial scales, both large and relatively
small, where cosmic vacuum dominates by density over dark matter
and baryons. The dynamical effect of cosmic vacuum is enhanced by
the fact that the effective gravitating density of vacuum is
$\rho_V + 3p_V = -2 \rho_V$, where $p_V = - \rho_V$ is the vacuum
pressure. In the Local Volume, vacuum dominates by force at the
distances larger than $R_V \simeq$ 1.5--2 Mpc from the barycenter
of the Local Group. On the `zero-gravity surface' of the size
$\simeq R_V$, the gravity of the Local Group dark matter and
baryons is balanced by the antigravity of vacuum. Observations
show that these are just the distances from which the observed
Hubble flow takes start.

These considerations suggest that cosmic vacuum may control the
dynamics of the observed Universe at both global spatial scales
approaching the observation horizon and local scales deep inside
the cell of matter uniformity. Because of this, the cosmological
expansion may be not only a global phenomenon, but also a local
one. On all the spatial scales where vacuum dominates, the
expansion rate may be exactly or nearly the same, since the rate
(measured by the Hubble factor) is mainly determined by one and
the same physical agent which is vacuum with its perfectly uniform
density (Karachentsev, Chernin, Teerikorpi 2003).

As Sandage and his colleagues comment this idea, ``...the total
force field is nearly homogeneous (smooth) due to the dominance
of an all pervasive cosmological constant, diluting any lumpy
gravity field of the clustered matter...'' (Thim et al. 2003).

Cosmological N-body LambdaCDM simulations performed by Macci\`{o}
et al. (2005) are reported to confirm both qualitative and
quantitative conclusions of our papers mentioned above. A theory
treatment by Leong and Saslaw (2004) within the framework of
many-body gravitational clustering shows that the observed small
departures from the regular Hubble flow beyond the Local Group
are highly probable. They also find that low mass groups, like
the Local Group, dominate, and the Hubble flow is not
significantly disturbed around them. Important observed features
of the local expansion flow are discussed by Whiting (2003),
Axenides and Perivolaropoulos (2002).

The nearest to us part of the expansion flow, the very local ($<$
3 Mpc) Hubble flow (VLHF), is obviously of a special interest: it
is the area where the expansion takes its origin. In Paper I, we
presented most recent data on the VLHF kinematics including
original distance measurements made mostly with the {\em Hubble
Space Telescope} (Karachentsev et al. 2000, 2002, 2003). The data
display the VLHF as a well organized outflow with the linear
velocity-distance relation. Its expansion rate is $72 \pm 15$ km
s$^{-1}$ Mpc$^{-1}$ and the one-dimension random mean motion is
about 30 km s$^{-1}$.

As a first step in the search for the VLHF physical nature, we
performed a set of computer simulations to reconstruct the
dynamical history of the flow. We found (Paper I) that the
evolution of the VLHF might begin at the epoch of the Local Group
formation some 12.5 Gyr ago. At that time, the VLHF galaxies,
together with the forming major galaxies of the group and a
variety of subgalactic units, participated in violent nonlinear
dynamics with multiple collisions and merging. This violent
initial state is essentially similar to the Little Bang model
(Byrd et al. 1994) and the concept of the early Local Group
described by van den Bergh (2003). We demonstrated that the VLHF
might be formed by relatively small units that survived accretion
by the major galaxies and managed to escape from the gravitational
potential well of the Local Group to the area outside the local
zero-gravity surface. A typical VLHF member galaxy gained escape
velocity from the non-stationary gravity field of the forming
group and a velocity larger than some 200~km~s$^{-1}$ enabled it
to reach the vacuum-dominated area where they gained additional
acceleration from cosmic vacuum.

According to the results of Paper I, the VLHF is not a slightly
disturbed initial cosmological flow with the linear velocity law
existed from the `very beginning'. On the contrary, the observed
members of the VLHF underwent strong nonlinear interactions with
the major galaxies of the Local Group before they formed the
quasi-regular flow outside the zero-gravity surface. Their
initially chaotic motions became quasi-regular under the combine
action of the quasi-static and quasi-spherically symmetric
gravitational potential of the Local Group outside the local
zero-gravity surface and the vacuum which controlled the flow
further dynamics there.

In the present paper, we develop the systematic studies of the
VLHF along the line traced in Paper I. We focus now on the
physics which controls the flow transition from the initial chaos
to the present-day order. Two key aspects of the problem are the
dynamical cooling of the expansion flow outside the local
zero-gravity surface, which is studied analytically in Sec.2, and
the structure of the chaotic motions inside this surface, which
is a subject of computer simulations presented in Sec.3. A
discussion of the results is given and the major conclusions are
summarized in Sec.4.

\section{Vacuum cooling}

Following the results of Paper I, we assume that the VLHF takes
its origin in the Little Bang developed inside the zero-gravity
(ZG) surface. Each VLHF galaxy participates in the violent
nonlinear dynamics of a many-body system of primeval galaxies and
gaseous subgalactic clumps moving in the non-stationary
gravitational potential of the early Local Group some 12.5 Gyrs.
When a galaxy crosses the ZG surface, it enters the area where
the gravitational potential is nearly spherically symmetric and
almost static. The former is clearly indicated by the shape of
the local ZG surface which is nearly spherical and only slightly
changes approaching in shape to the perfect static sphere,
asymptotically (Figs.5--8 below; see also Dolgachev et al. 2003,
2004 for more details). A typical trajectory of the small body (a
dwarf galaxy) is nearly radial, in such a potential, as computer
simulations show (see Sec.3). Therefore the first stage of the
transition from chaos to order occurs within the ZG surface, when
an initially chaotic outflow of escaping galaxies becomes more or
less radial reaching the sphere. But this nearly radial flow
remains rather `hot': its radial velocity dispersion is still high
at the surface. The outflow cooling is the second stage of the
transition process.

In this section, we focus on the second stage of the transition
process. Due to the nearly radial symmetry of the outflow, a
simple analytical treatment is possible for this stage (the first
stage is studied in the next section with the use of computer
simulations).

Let us follow the motion of a small body (a test particle)
outside the ZG surface along a radial trajectory. We will consider
the particle a `typical' member of the VLHF. Neglecting in the
first approximation any (small) deviations of the `matter part'
of gravitational potential from the spherical shape, we may
describe the trajectory by the Newtonian equation of motion in the
form:
% 1
\be \ddot r(t) = - GM/r^2 + r/A_V^2, \;\;\; r \ge R_V. \ee

The gravity of matter and the antigravity of vacuum are
represented in the right-hand side of the equation by the terms
of the opposite signs. Here $r$ is the distance of the body to the
Local Group barycenter, $M = 2 \times 10^{12} M_{\odot}$ is the
mass of the group, including both dark matter and baryons. The
value $A_V = (\frac{8 \pi G}{3} \rho_V)^{-1/2} \simeq 1.5 \times
10^{28}$ cm is the vacuum characteristic length (the Friedmann
integral for vacuum --- see Chernin 2001), and $\rho_V = 7 \times
10^{-30} $ g cm$^{-3}$ is the vacuum density. (Hereafter the
speed of light $c = 1$ in the formulas.)

The ZG surface at which $\ddot r = 0$ is a sphere of the radius
% 2
\be R_V = (GM A_V^2)^{1/3} \simeq 1.5 \; {\rm Mpc.} \ee

The mathematical structure of Eq.1 is similar to the structure of
the Friedmann cosmological equation for the global expansion
scale factor $R(t)$:
% 3
\be \ddot R (t, \chi) = - G M(\chi)/R^2 + R/A_V^2, \ee where $M
(\chi) = {\rm Const} (t)$ is the mass of non-relativistic dark matter
and baryons within the sphere of the Lagrangian radius
$R(\chi,t)$, and $\chi$ is the Lagrangian coordinate of the body.
The difference of Eq.3 from Eq.1 is in the dependence of $M$ on
the Lagrangian coordinate. According to Eq.1, each body moves in
the potential of the same gravitational mass, while Eq.3 indicates
that the mass is different for different $\chi$, i.e. different
particles. The `vacuum part' of the gravitational potential and
force is the same in both Eq.1 and Eq.3. Asymptotically, when $r$
and $R$ go to infinity, vacuum dominates entirely, and the
difference between the equations vanishes: they coincide
completely. In this limit, both equations give the same
velocity-distance law $\dot r = r/A_V$ and $\dot R = R/A_V$. Then
the expansion rate, the Hubble factor $H_V = 1/A_V$ is a constant
in both cases. This asymptotic similarity of the equations is
significant: it indicates that the motion of the particle we
follow tends asymptotically to the cosmological regime. In
conventional units, $H_V = 1/A_V \simeq 60$ km s$^{-1}$ Mpc$^{-1}$.
This value is in agreement with the observational data; according
to (Karachentsev et al. 2002, 2003), the expansion rate for
the VLHF is $72 \pm 15$ km s$^{-1}$ Mpc$^{-1}$ (see Sec.1).

A close relation to cosmology is seen not only in the vacuum part
of the potential, but also in its matter part. Indeed, the total
mass of the Local Group (which is mostly dark matter mass in the
halos of the two major galaxies of the group) together with the
mass of all the bodies of the VLHF are collected from the
initially uniform cosmological matter distribution. An
`unperturbed' cosmological volume containing the non-relativistic
mass $M$ has the present-day radius
% 4
\be R_M = \left[M /(\frac{4 \pi}{3} \rho_M)\right]^{1/3} \simeq 2
\; {\rm Mpc.} \ee
Here $\rho_M = \rho_D + \rho_B \simeq 3 \times 10^{-30}$ g
cm$^{-1}$ is the sum of the dark matter density and the baryonic
density at present. A near coincidence of the values $R_M$ and
$R_V$ is obviously a result of the near coincidence of the mean
dark matter density and the vacuum density in the present-day
state of the Universe.

This means that the analytical model described by Eq.1 has a
direct connection to cosmology, since its both parameters, $M$
and $A_V$, have a clear cosmological meaning: cosmology
parameters are imprinted in the dynamical background on which the
local expansion flow develops.

The first integral of Eq.1 expresses, as usual, the mechanical
energy conservation:
% 5
\be \frac{1}{2}\dot r^2 = GM/r + \frac{1}{2}(r/A_V)^2 + E, \ee
where $E = {\rm Const}$ is different, generally, for different
particles. When the trajectory crosses the ZG surface and $r =
R_V$, one has from Eq.5:
% 6
\be \dot r^2 = 3 (R_V/A_V)^2 + 2 E, \;\;\; r = R_V. \ee

Let us introduce a `random' velocity $v$ as a difference between
the velocity $\dot r$ and a `regular' (asymptotic) velocity
$r/A_V$. Then the random velocity $v$ at $r = R_V$,
% 7
\be v_1 = (R_V/A_V)\left[(3 + 2 EA_V^2/R_V^2)^{1/2} -1\right]. \ee

As the trajectory reaches the distance $r$ outside the ZG
surface, the random velocity $v(r)$ is given (for an arbitrary
$E$) by the relation:
% 8
\be v (r) = (R_V/A_V)\left[(r^2/R_V^2 - 2 + 2R_V/r + 2 v_1A_V/R_V +
v_1^2A_V^2/R_V^2)^{1/2} - r/R_V\right]. \ee

A comparison of $v(r)$ with $v_1$ enables to find how the random
velocity decreases along the particle trajectory. For this goal,
we may introduce the `vacuum cooling factor', $q_V \equiv v_1/v$
which measures the efficiency of the random velocity suppression
due to vacuum:
% 9
\be q_V = (v_1 A_V/R_V)\left[(r^2/R_V^2 +2 + 2R_V/r + 2 v_1A_V/R_V +
v_1^2A_V^2/R_V^2)^{1/2} - r/R_V\right]^{-1}. \ee

In the simplest case when the total energy is zero, $E = 0$, the
velocity $v_1 = (\sqrt{3} -1) R_V/A_V$, and we have from Eq.8:
% 10
\be v (r) = (R_V/A_V)\left[(r^2/R_V^2 +2 R_V/r)^{1/2} - r/R_V\right]. \ee
Then the vacuum cooling factor in the parabolic expansion flow
% 11
\be q_V (r) = v_1/v(r) = (\sqrt{3} - 1)\left[(r^2/R_V^2 +2)^{1/2}
- r/R_V\right]^{-1}. \ee

We may see from Eqs.7,11 that the random velocity is diminished by
factor $\simeq 3$ --- from  $\simeq 70$ to $\simeq 20$ km s$^{-1}$ ---
during the time when the body covers the path from $r = R_V$ to
the VLHF upper spatial limit $r \simeq 2 R_V \simeq 3$ Mpc. The
resulting random velocity is in agreement with the observational
value of 30 km s$^{-1}$ (see Karachentsev et al. 2002, 2003 and Sec.1).
When the same body reaches, say, distances 4 or 6 $R_V$, the
cooling factor increases to $q_V = 12$ and $q_V = 56$,
correspondingly.

If the random velocity $v(r)$ is small compared to the regular
velocity $r/A_V$ and $E = 0$, one finds from Eq.5 in the linear
approximation:
% 12
\be v \simeq R_V^3/(A_V r^2) \propto r^{-2}. \ee

Recall that the cosmological adiabatic cooling is described by
the relation $v \propto R^{-1}$. We may conclude that the vacuum
cooling of the local Hubble flow acts much more effectively than
the adiabatic cooling.

The second integral of Eq.5 (for $E=0$) has a well-known exact
solution: $r \propto \sinh[\frac{3}{2}t/A_V]^{2/3}$. Then the law
of vacuum cooling in the linear approximation takes the form:
% 13
\be q_V \propto\sinh[\frac{3}{2}t/A_V]^{4/3}. \ee

It is interesting --- for comparison --- to follow the dynamics of
the same expansion flow, but in the absence of cosmic vacuum. With
the same statement of the problem as above, but with $\rho_V = 0$
and $E >0$, we would have an asymptotic (as $r$ goes to infinity)
motion which is an inertial one: $\dot r = (2 E)^{1/2}$. This is
also a Hubble-type motion with the linear velocity-distance
relation: $\dot r = r/t$, where the expansion rate, $H_E = 1/t$,
depends on time. This suggests that the flow may be cooled without
vacuum as well (see also Sec.3); but the cooling efficiency is
much lower in the absence of vacuum. Indeed, taking $r/t$ as a
regular motion velocity, we may introduce the random velocity $v$
by the relation $\dot r = r/t + v$. Then from Eq.5 (without the
vacuum term) we find:
% 14
\be v = (2E)^{1/2} \left[(1 + \frac{GM}{2E r})^{1/2} -1\right]. \ee
In the linear approximation for $v/(2E)^{1/2} < 1$, this leads to the
adiabatic relation $v \propto r^{-1}$. As a result, the cooling
factor in the absence of vacuum, $q_E \propto r$, is significantly
smaller than the vacuum cooling factor $q_V$ of Eqs.9, 11, 13.

Thus, vacuum cooling reveals a high efficiency resulting in the
rapid suppression of the random velocity in the outflow outside
the local ZG surface. The quantitative similarity of the
resulting VLHF parameters is due to the fact that the outflow
dynamics develops on the dynamical background with the
cosmology-related parameters.

\section{Computer simulations: from Little Bang to VLHF}

The VLHF evolution inside the local ZG surface cannot be
described in any simple analytical model. It is controlled by
complex nonlinear dynamics which may be followed --- but only to
some extent --- with computer simulations. The Little Bang model
(Byrd et al. 1994) and the picture of the early Local Group (van
den Bergh 2003) provide important insights to this dynamics.
Basing on this, we developed a simplified approach in which the
free fall of the two major galaxies of the Local Group on the
cosmic vacuum background plays a central part. This is a `Little
Bang Minimal Model' (LBMM) which takes into account that 1) the
mass of the Local Group (LG) is mostly dark mass and it strongly
concentrates (Karachentsev et al. 2002) to the two major galaxies
of the group; 2) the dark matter halos of the Milky Way (MW) and
the Andromeda Galaxy (AG) are nearly spherical; 3) the relative
MW-AG motion is directed along the line of the centers of the
galaxies, in accordance with the classic treatment (Kahn and
Woltjer 1959); 4) the LG is embedded in cosmic vacuum which is
represented by perfect medium with a uniform constant energy
density.

The parameters of the model are the total mass of MW
$1\times10^{12} M_{\sun}$, the total mass of AG $1.5\times10^{12}
M_{\sun}$ and the vacuum density $\rho_V = 7 \times 10^{-30}$
g/cm$^3$ (so the total mass of the Local Group is somewhat larger
here than in Sec.2). With the present separation 0.7 Mpc and the
present relative velocity $-120$~km~s$^{-1}$, the two major
galaxies of the LG started their motion toward each other 12.5
Gyr ago, in the LBMM. Like in the section above, the motions of
the VLHF dwarf galaxies inside are considered as test particle
motions, so that the problem under consideration is reduced to
the three-body restricted problem. Despite its obvious
simplicity, the LBMM demonstrates a variety of dynamical patterns
that might play, as we demonstrate below, the central part in the
origin and evolution of the VLHF.

The initial (12.5 Gyr ago) conditions for the particle
trajectories are chosen in the three-dimensional axially
symmetrical phase space of positions and radial velocities of the
LBMM. The phase space is scanned uniformly in the radial velocity
range from $-500$ to 500 km s$^{-1}$ and the radial distance range from
0.15 to 0.80 Mpc. Both velocities and distances are related to the
LG center-of-mass. Two sets of 420 trajectories each are
integrated --- one in the model with cosmic vacuum and the other
in the model with no vacuum, for comparison. According to the
results of the integration, 397 and 409 trajectories of the first
and the second sets, respectively, are found in the VLHF distance
interval (from 1.5 to 3 Mpc) at the present-day epoch. These two
subsets serve as the banks of trajectories for the further
statistical analysis. To simulate the observed sample of the 22
VLHF galaxies (its description see in Paper I), we used model
samples of 22 trajectories each selected randomly from each of
the two subsets of trajectories.

Table 1 shows 10 such simulation samples of trajectories in the
model with the observed vacuum density. Each sample is
characterized by the linear regression factor (calculated with
the least square method) and the corresponding linear velocity
dispersion. The both are calculated for the present and the
initial states of the sample. The present regression factor (the
first line in Table 1) gives the expansion time rate which may be
compared with the observed VLHF Hubble factor. The present
velocity dispersion (the second line) is to be compared with the
observed velocity dispersion. The initial regression factor (the
third line) and the initial velocity dispersion (the fourth line)
give a quantitative measure of the chaotic state of the flow 12.5
Gyr ago.

\vspace{3cm}
%\documentstyle [11pt]{article}
%\setlength{\textheight}{21.6cm} $\setlength{\textwidth}{14.7cm}
%\setlength{\baselineskip}{13pt} $\setlength{\topmargin}{0cm}
%\setlength{\oddsidemargin}{0.4cm}
%\setlength{\evensidemargin}{0.4cm} $\begin{document} \noindent
%\newpage
Table 1 \\[5pt]
%\smallskip
\begin{tabular}{|c|c|c|c|c|c|c|c|c|c|}
\hline
   1   & 2   &  3   &  4   &  5    &  6   &   7   &   8   &  9   &10 \\[2pt]
\hline
       &       &      &      &       &      &       &       &      &        \\[-7pt]
  86.30&  98.85& 69.41&145.20&  91.95&123.08& 104.24& 100.82&115.05& 74.91  \\[2pt]
  30.29&  31.46& 32.53& 27.16&  36.81& 31.29&  20.93&  29.52& 36.80& 34.66  \\[-7pt]
       &       &      &      &       &      &       &       &      &        \\
$-225.96$&$-109.43$&$-74.81$&311.66&$-121.96$&278.96&$-130.24$&$-117.33$&$-12.17$&$-152.04$ \\[2pt]
 185.13& 198.54&182.22&193.88& 193.20&187.50& 197.39& 171.62&178.48& 198.94\\[-7pt]
       &       &      &      &       &      &       &       &      &        \\
\hline
\end{tabular}

\vspace{9mm}
Table 2 \\[5pt]
\begin{tabular}{|c|c|c|c|c|c|c|c|c|c|}
\hline
   1   & 2   &  3   &  4   &  5    &  6   &   7   &   8   &  9   &10 \\[2pt]
\hline
       &      &      &      &      &      &       &      &       &        \\[-7pt]
  74.93& 60.55& 86.66& 85.87&105.07& 97.17&  87.03& 56.57&  78.80& 46.80  \\[2pt]
  36.99& 37.50& 29.59& 39.59& 37.15& 32.15&  34.98& 36.91&  34.22& 33.49  \\[-7pt]
       &      &      &      &      &      &       &      &       &        \\
 $-19.39$&153.65&123.25&301.90&362.22&$-3.34$&$-173.21$&$-25.99$&$-331.72$&  80.69 \\[2pt]
 198.97&185.39&198.75&192.67&174.79&203.86& 209.54&212.12& 201.32& 203.77\\[-7pt]
       &      &      &      &      &      &       &      &       &        \\
\hline
\end{tabular}

\vspace{9mm}
Table 3 \\[5pt]
\begin{tabular}{|c|c|c|c|}
\hline
   1   &$Vr\pm50$&$Vr\pm100$&$0\pm100$ \\[2pt]
\hline
       &         &          &        \\[-7pt]
  86.30&    84.47& 78.02    & 27.56  \\[2pt]
  30.29&    28.45& 24.60    & 65.27  \\[-7pt]
       &         &          &        \\
\hline
\end{tabular}

\vspace{2cm}

Let us look, for instance, at the figures of Sample 1 in Table 1.
The present-day state of the sample imitates well  the observed
VLHF. Indeed, the regression factor is 86 km s$^{-1}$ Mpc$^{-1}$
which is
within the one-sigma interval of the observed value of the local
Hubble parameter, $H_L = 72 \pm 15$ km s$^{-1}$ Mpc$^{-1}$
(Karachentsev et al. 2002, 2003).
The flow is quiet and cool: its velocity dispersion
is 30 km s$^{-1}$ which coincides exactly with the observed value. The
present-day Hubble diagram for Sample 1 is showed in Fig.1; it
looks very similar to the the real Hubble diagram for the VLHF
(Karachentsev  2001).

The initial state of the sample differs drastically from the
present one. Indeed, the initial regression factor is negative
for Sample 1 of Table 1; it means that the flow was contracting,
not expanding, 12.5 Gyr ago. The corresponding initial velocity
dispersion is rather high, near 200 km s$^{-1}$. Both figures show
clearly that the initial state was highly chaotic and had nothing
in common with an imaginary regular (or slightly disturbed)
cosmological expansion flow on similar spatial scales some 12.5
Gyr ago.

Looking at the other samples of Table 1, we may see that, at the
present-day state of the model samples, they are characterized by
the regression factors which spread from 74 to 123
km s$^{-1}$ Mpc$^{-1}$, and
by the velocity dispersion within a rather narrow interval 27--37
km s$^{-1}$. Their initial states are highly chaotic and may be both
contracting or expanding. The initial velocity dispersion is near
200 km s$^{-1}$ for all the samples.

Table 1 and Fig.1 demonstrate that the LBMM phase space may host
highly chaotic sets of the initial states, and they give rise
eventually quasi-regular outflows with the linear
velocity-distance relation and the expansion rate close to the
observed one. It is most remarkable that the present velocity
dispersion values are confined within a rather narrow interval,
between 20 and 40 km s$^{-1}$.

The internal dynamical structure of the samples of Table 1 in the
present-day state is illustrated by histograms of Fig.2. The
histograms give the distributions of the `individual expansion
rates' which are velocity-distance ratios for the members of a
sample. We may see that, in a typical case, there is a dominant
group of trajectories which determines the overall statistical
characteristics of the samples.

Table 2 shows the data for the simulation samples in the model
without vacuum. We may see that the major trend in the dynamical
evolution is the same: this is a transformation of a flow from
chaos to order. The similarity with the vacuum model in this
respect is not quite unexpected, in view of the analysis of
Sec.2. Indeed, the existence of a regular asymptotic states with
the linear velocity-distance relation is rather universal, as is
seen from both analytical treatment and computer simulations
described by Figs.1,2. The vacuum effect enhances the trend tto
order essentially, and this reveals clearer at larger distances in
the local expansion flows. On the modest distances of the VLHF,
the quantitative difference between the two models is not too
significant. One may only notice somewhat lower values of the
final regression factor, between 47 and 105 km s$^{-1}$ Mpc$^{-1}$,
and somewhat
larger values of the velocity dispersion in a narrow range from
32 to 38 km s$^{-1}$, in the models without vacuum.

Taking Sample 1 of Table 1 as a typical one, we show its 22
trajectories in the plane of symmetry of the Local Group --- see
Fig.3. The local ZG surface is showed in the section of this
plane; both initial (dash line) and present day (solid line)
shape of the surface are given in the figure. It is seen that the
surface is rather circular and change with time only very
slightly for the last 12.5 Gyr. The most of the trajectories
intersect the ZG surface along nearly radial directions. The
geometry of most of the trajectories is rather simple: they are
very close to straight lines, outside the ZG surface (this fact
was used in the analytical treatment of Sec.2). At least in part,
such a geometry is due to the initial conditions at which the
velocities had only a radial component.

A possible effect of initial non-radial velocities was especially
studied for same Sample 1. For this goal, the dimensionality of
the LBMM initial phase space was extended from 3 to 4, to include
transverse velocities. Fig.4 shows the trajectories of the same
galaxies, but with the additional transverse velocity which is
initially $\pm 50$ km s$^{-1}$ for the members of the sample. Fig.5
illustrates the effect of initial transverse velocities of $\pm
100$ km s$^{-1}$. Finally, Fig.6 shows  (unrealistic) trajectories with
zero initial radial velocity and with initial transverse
velocities of $\pm 100$ km s$^{-1}$; this is actually a case of an
initial rapid rotation of the masses of the ensemble around the LG
barycenter, and most of the trajectories prove to be kept within
the local ZG surface. With the exception of the unrealistic
example of Fig.6, the initial transverse velocities do not alter
significantly the simple nearly straight-line geometry of typical
trajectories outside the ZG surface.

In addition, Table 3 gives quantitative characteristics of the
effect of the transverse velocities --- the present expansion rate
(the first line) and velocity dispersion. A comparison with the
data for Sample 1 of Table 1 shows that the realistic initial
transverse velocities (columns 2,3) are compatible with the major
trend of the sample evolution. They do not change practically the
statistical characteristics of the outflow present-day state.
Indeed, with the transverse velocities $\pm 50$ and $\pm 100$
km s$^{-1}$, the expansion rate decreases slightly (from 86 to 85 and 78
km s$^{-1}$, correspondingly) keeping within the same one-sigma interval
of the observation figure. The velocity dispersion does not change
significantly either. Only in the unrealistic case of zero initial
radial velocities (column 4), the figures turn out to be
considerably different.

Thus, almost a thousand model trajectories have enabled us to
simulate the origin of the VLHF and its dynamical evolution
within the zero-gravity surface and out of it. In agreement with
the analytic considerations of Sec.2, the simulations prove the
real possibility of the VLHF evolution from the Little Bang chaos
to the order and regularity at present.

\section{Discussion and conclusions}

The recent high precision mapping of the local velocity field and
high accuracy distance measurements in the Local Volume
(Karachentsev et al. 2000, 2001, 2002, 2003, 2004; Karachentsev
and Makarov 2000; Ekholm et al. 2001; Davidge and van den Bergh
2001; Teerirorpi and Paturel 2002; Thim et al. 2003; Paturel and
Teerikorpi 2004, Reindl et al. 2005) have given a clear picture of
the nearest Universe. The results have reliably confirmed the
earlier conclusions by Sandage et al. (1972) and Sandage (1986,
1999): there is a regular local (within 20 Mpc distance) Hubble
flow of expansion deep inside the cell of matter uniformity; it
has the expansion rate near the global Hubble constant and the
remarkably low velocity dispersion. An understanding of the
physical nature of the local Hubble flow has come with the
discovery of cosmic vacuum (or the cosmological constant): vacuum
with its  perfectly uniform dominant density provides dynamical
conditions for a regular flow in the highly non-uniform matter
distribution (Chernin et al. 2000, Chernin 2001, Baryshev et al.
2001, Karachentsev et al. 2003).

Our further attempts along this line of studies are focused on
the very local ($\le 3$ Mpc) area of the flow where it takes
start. It has led us to a conclusion (Paper I) that the very
local Hubble flow was generated in the Little Bang due to the
chaotic dynamics of the forming Local Group. In the present
paper, we studied the physics of the transition from the initial
chaotic state of the flow to its presently observed regular state.

The transition from chaos to order is known in various areas of
natural (and also social) sciences, but poorly understood yet. It
has been found in fundamental nonlinear physics and, in
particular, in the theory of turbulence, that a high degree of
complexity is characteristic for this process. Remarkably enough,
in the case of the very local Hubble flow (VLHF), the nonlinear
phenomena we face may be clarified and treated in a fairly simple
way. The Little Bang Minimal Model (LBMM) enables us to describe
the problem in an explicit form and find both qualitative and
quantitative solutions to it. The solution we obtain proves to be
in a good agreement with the high precision observations
mentioned above.

Indeed, the nature of the VLHF initial chaos is not too puzzling
in our model. It is, first, due to random choice of the initial
states of the VLHF members in the phase space of the Local Group
embedded in cosmic vacuum. Second, it is due to the nonlinear
three-body dynamics of gravitational scattering of a test mass on
a non-stationary massive binary system. The gravitational
scattering is a complex process; the dynamical instability is the
major physical factor that controls it and introduces the 
Poincar\'{e}-type chaos to the system (see, for a reviews Valtonen and
Mikkola 1991, Chernin and Valtonen 1998). But numerical solutions
and computer simulations of this process may easily be performed:
under the particular conditions described by the LBMM, the
simulations are reduced to the standard integration of the
three-body restricted problem (Sec.3). The only new feature is
the vacuum background on which the process develops.

It turns out that, when the test body escapes from the two-body
gravitational potential, the dynamical instability gradually
ceases with the distance from the system barycenter, and the
third body occurs in the potential which becomes more and more
centrally symmetrical. This is clearly seen from the computed
geometry of the local zero-gravity (ZG) surface. In fact, the
escaped body is affected mainly by a simple nearly
centrally-symmetrical potential. This potential is practically
static: its evolution is very slow, as is seen from the
comparison of the initial and final shape and size of the
zero-gravity surface (Figs.3--6). Under such conditions, the motion
of the scattered body becomes more and more regular. In a sample
of bodies imitating the VLHF, it leads to a picture of a flow
which acquires the linear velocity-distance relation, and the
deviations from this relation decrease with time.

Cosmic vacuum which starts to dominate dynamically outside the ZG
surface of the Local Group introduces additional stability and
regularity to the sample evolution. First, it produces perfectly
symmetrical and static major contribution to the gravitational
potential. Second, it accelerates the escaped bodies and so leads
to larger distances from the system barycenter for the life time of the
system. At large distances from the center, where vacuum dominates
completely, the velocity-distance ratio tends to a constant value
$H_V \simeq 60$ km s$^{-1}$ Mpc$^{-1}$,
independently of the initial conditions.

Our computer simulations enable us also to find that, at the small
and modest distances characteristic for the VLHF galaxies, the
dynamical tendency introduced by cosmic vacuum is quite
recognizable, while it is not too significant quantitatively. But
at larger distances like 4--8 Mpc in the local Hubble flow, the
dynamical effect of vacuum is most important both qualitatively
and quantitatively.

Outside the ZG surface, the motions of the VLHF members can also
be studied with the use of exact analytical solutions. The
solutions give a description of the flow evolution to the
quasi-regular final state. The solutions show that the flow reach
asymptotically a regular structure with the linear
velocity-distance relation. This is found in the solutions with
vacuum and even in the solutions with no vacuum (Sec.2). In both
cases, the flow motion  has an asymptotic regime, which is
described by a simple, stable and self-similar solution. In
accordance with a general result of nonlinear mechanics, this
solution is a dynamical attractor for wide classes of more complex
solutions.

However the asymptotic regular regime is reached with different
efficiency, in the two cases under consideration. The most
efficient evolution from chaos to order proceeds in the solution
with cosmic vacuum. A quantitative measure of the efficiency of
this evolution is the `vacuum cooling factor' introduced and
estimated in Sec.2. It describes the vacuum contribution to the
VLHF evolution and demonstrates --- together with the computer
simulations --- that cosmic vacuum provides the most effective
mechanism of the flow transition from chaos to order, especially
at larger distances in the local Hubble flow. It is cosmic vacuum
that makes the whole local flow of expansion --- from the ZG
surface to its maximal spatial scales --- a self-consistent regular
cosmological phenomenon.

\begin{acknowledgements}
We thank Yu. Efremov, G. Byrd, A. Silbergleit, A. Cherepashchuk
and A. Zasov for discussions. DIM and ADC acknowledge support from
INTAS grant 03--55--1754 and the Russian Foundation for Basic
Research grant 03--02--16288. DIM is thankful to the Russian
Science Support Foundation for support.
\end{acknowledgements}

\newpage

\begin{figure}
\centerline{\psfig{figure=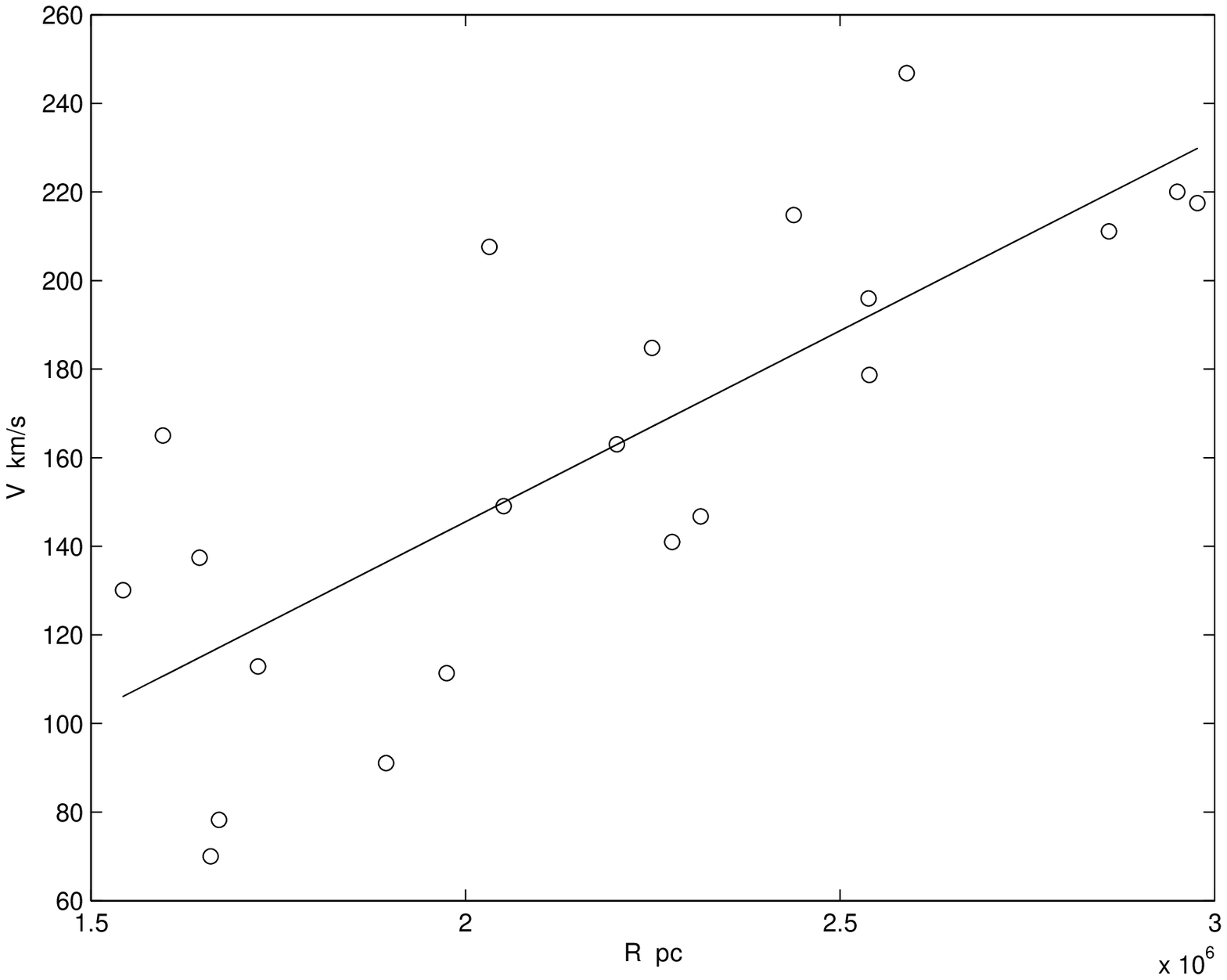,width=\textwidth}}
\caption{Velocity-distance diagram for Sample 1.\label{fig_1.eps}}
\end{figure}

\begin{figure}
\centerline{\psfig{figure=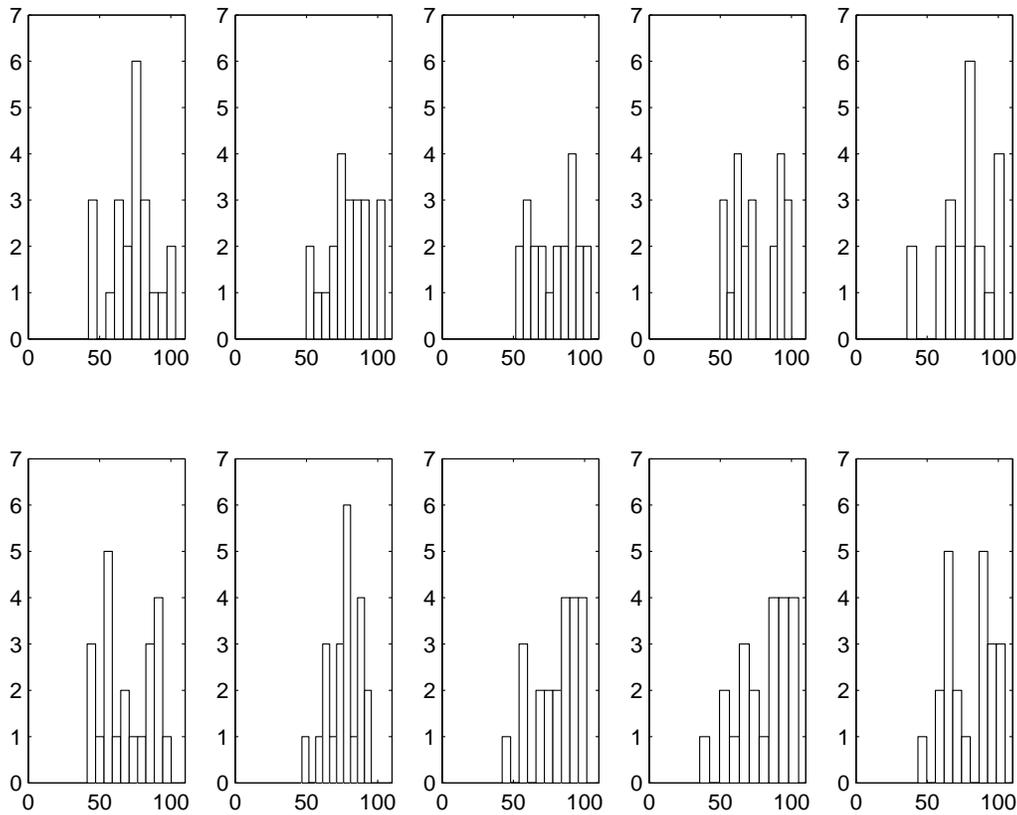,width=\textwidth}}
\caption{Internal kinematic structure of simulation samples.\label{fig1.eps}}
\end{figure}

\begin{figure}
\centerline{\psfig{figure=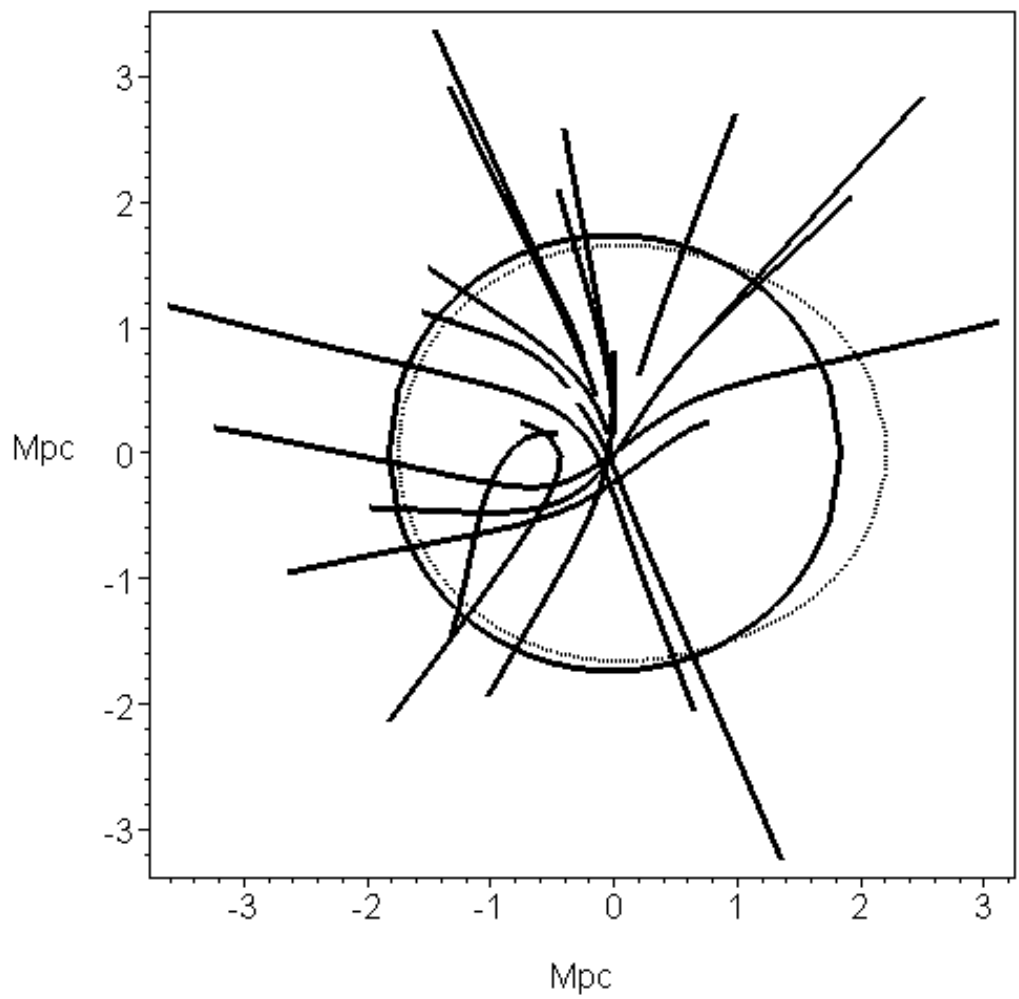,width=\textwidth,angle=0}}
\caption{Trajectories of Sample 1.\label{fig3.eps}}
\end{figure}

\begin{figure}
\centerline{\psfig{figure=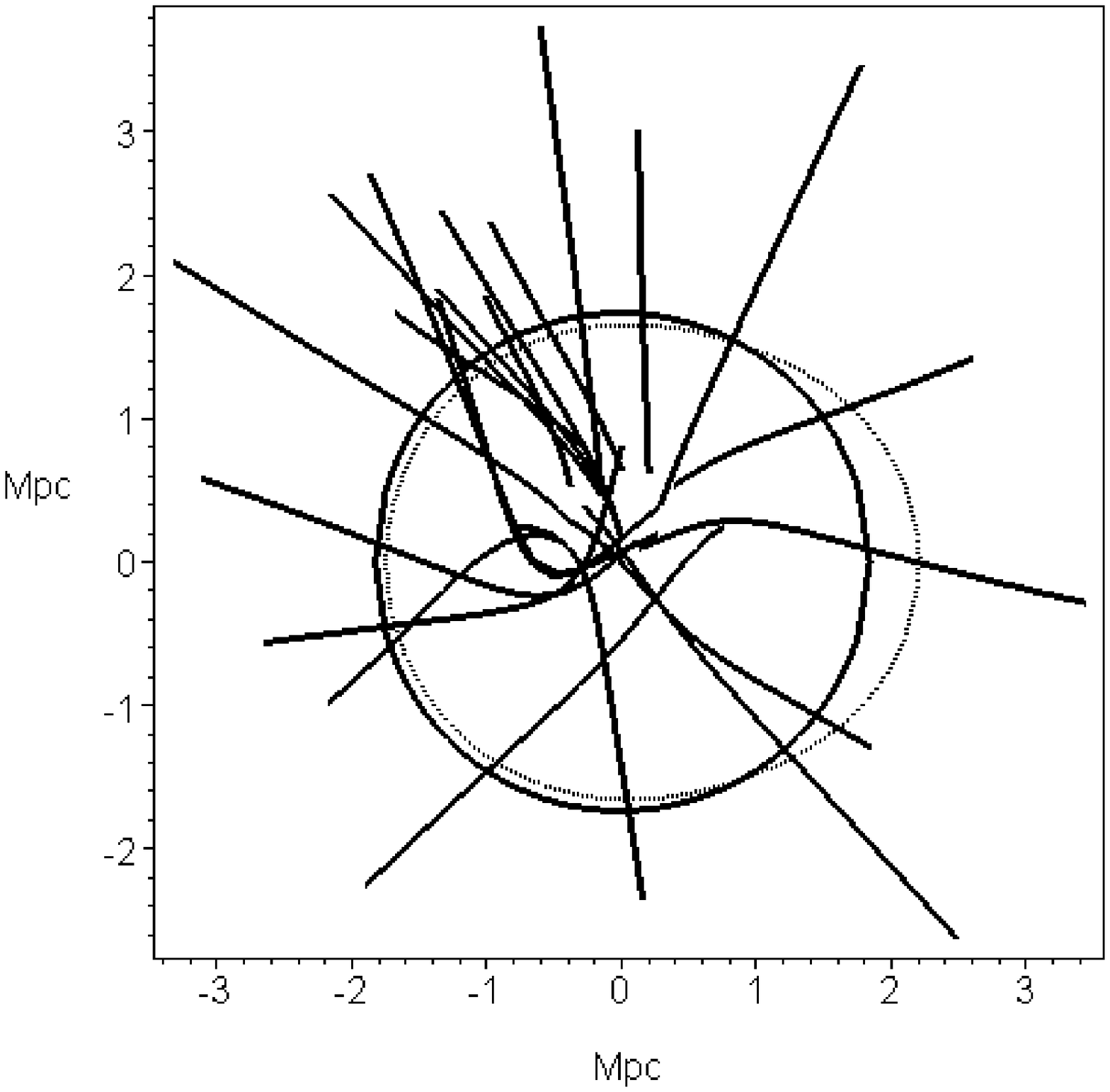,width=\textwidth,angle=0}} \caption{Same
for additional $\pm 50$ km s$^{-1}$ initial transverse
velocities.\label{fig4.eps}}
\end{figure}

\begin{figure}
\centerline{\psfig{figure=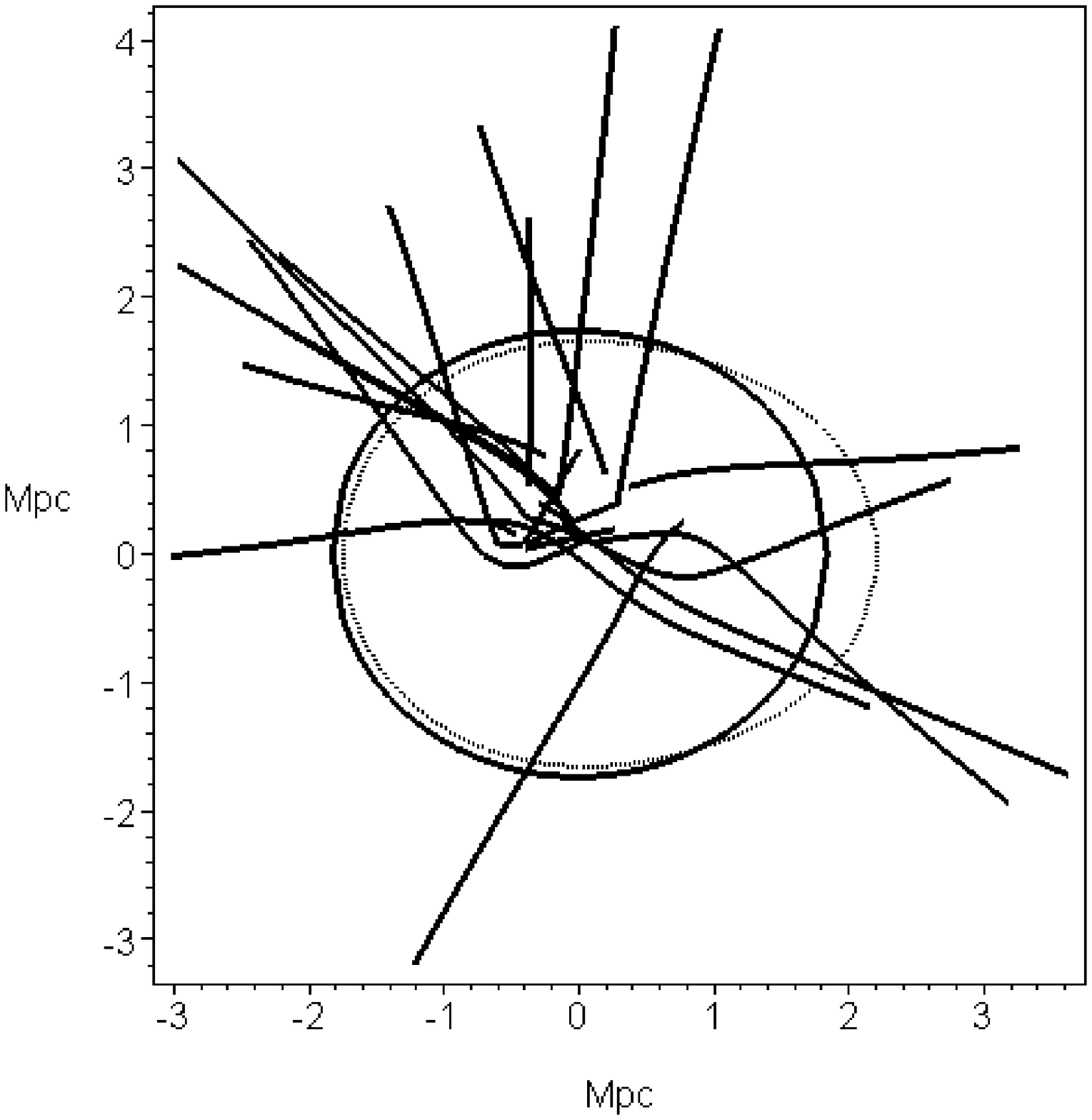,width=\textwidth,angle=0}} \caption{Same
for additional $\pm 100$ km s$^{-1}$ initial transverse
velocities.\label{fig5.eps}}
\end{figure}

\begin{figure}
\centerline{\psfig{figure=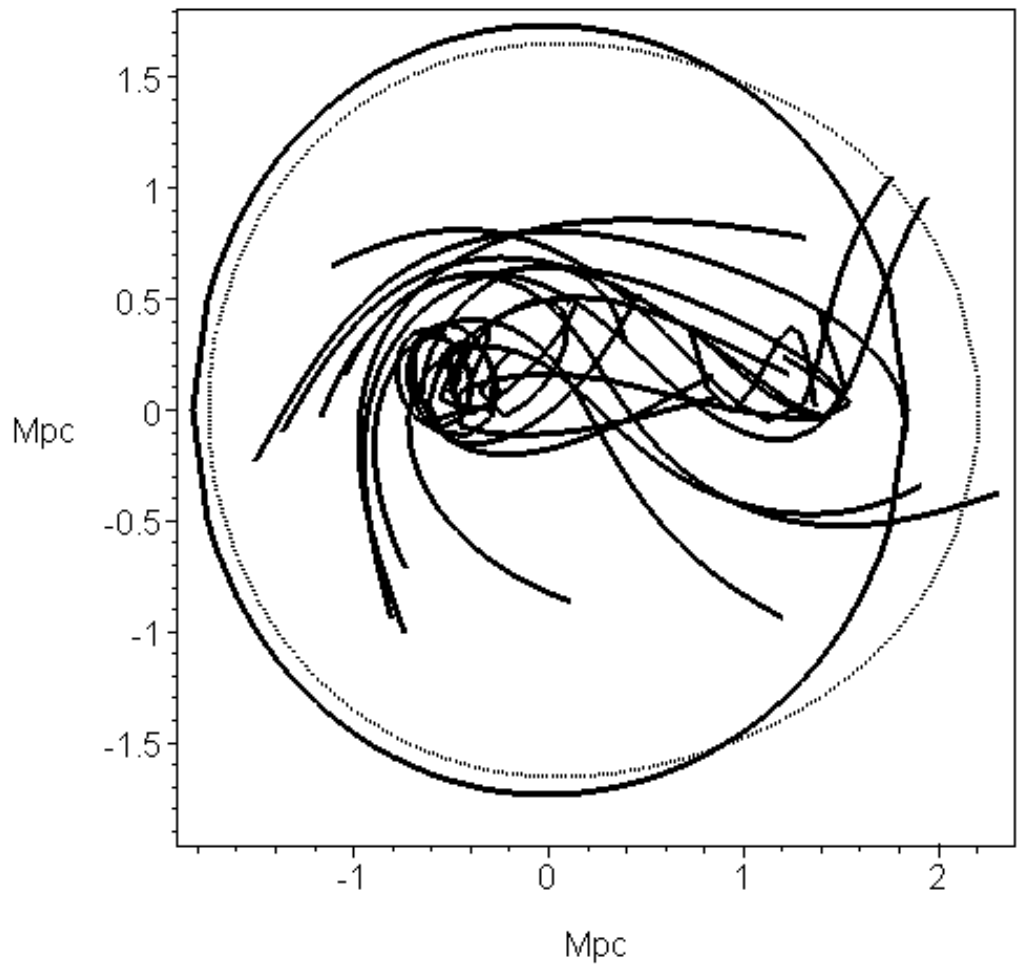,width=\textwidth,angle=0}} \caption{Same
for $\pm 100$ km s$^{-1}$ initial transverse velocities and no radial
velocities.\label{fig6.eps}}
\end{figure}

\end{document}